\documentclass[showpacs,showkeys,preprint]{revtex4}
\usepackage{latexsym}
\usepackage{amsmath}
\usepackage{amssymb}
\usepackage{amsthm}
\usepackage{graphicx}
\usepackage{dcolumn}
\newcommand{\be}{\begin{equation}}
\newcommand{\ee}{\end{equation}}
\newcommand{\ba}{\begin{eqnarray}}
\newcommand{\ea}{\end{eqnarray}}

\begin{document}
\title{Anomalous phase behavior of a soft-repulsive potential \\
with a strictly monotonic force}
\author{Franz Saija$^1$~\cite{aff1}, Santi Prestipino$^2$~\cite{aff2}
and Gianpietro Malescio$^2$~\cite{aff3}}
\affiliation{
$^1$ CNR-Istituto per i Processi Chimico-Fisici, Contrada Papardo,
98158 Messina, Italy \\
$^2$ Universit\`a degli Studi di Messina, Dipartimento di Fisica,
Contrada Papardo, 98166 Messina, Italy}
\date{\today}
\begin{abstract}
We study the phase behavior of a classical system of particles
interacting through a strictly convex soft-repulsive potential which,
at variance with the pairwise softened repulsions considered so far in the
literature, lacks a region of downward or zero curvature. 
Nonetheless, such interaction is characterized by two length scales, 
owing to the presence of a range of interparticle distances where the
repulsive force increases, for decreasing distance, much more slowly than
in the adjacent regions. 
We investigate, using extensive Monte Carlo simulations combined
with accurate free-energy calculations, the phase diagram of the system
under consideration.
We find that the model exhibits a  fluid-solid coexistence line with
multiple re-entrant regions, an extremely rich solid polymorphism with
solid-solid transitions, and water-like anomalies.
In spite of the isotropic nature of the interparticle potential, we find that,
among the crystal structures in which the system 
can exist, there are also a number of non-Bravais lattices, such as cI16 and diamond.
\end{abstract}
\maketitle

\section{Introduction}

In typical models of effective pairwise interaction between
particles, the repulsive force steadily increases, with an ever-increasing
rate, as particles get more and more closer to each other. This behavior,
which is typical of e.g. the Lennard-Jones potential, is originated by the
harsh quantum repulsion that is associated with the overlapping of
electronic shells. However, effective classical interactions where the
repulsive component undergoes some form of softening in a range of
interparticle distances are relevant for many physical systems.
Most of them belong to the realm of so-called soft matter
(solutions of star polymers, colloidal dispersions, microgels,
etc.)~\cite{Likos1,Likos2}, but also the adiabatic interaction potential
(electronic in origin) of simple atomic systems under high pressure is
manifestly of the soft-core type~\cite{Malescio,Prestipino}.

The effects of particle-core softening on thermodynamic behavior were
first investigated by Hemmer and Stell~\cite{Hemmer}, who analysed the
possible occurrence of multiple critical points and isostructural
solid-solid transitions. These authors considered pair potentials
with a hard core augmented with a finite repulsive component in the form
of a square shoulder or a linear ramp, features that may be pertinent
to some pure metallic systems, metallic mixtures, electrolytes, and
colloidal systems.
Few years later, Young and Alder~\cite{Young2} showed that the hard core
plus square shoulder potential gives origin to an anomalous melting line
similar to that observed in Cs or Ce. Later, Debenedetti~\cite{Debenedetti}
showed that systems of particles interacting via potentials whose repulsive
core is softened by a curvature change are capable of contracting when
heated isobarically (a behavior known as ``density anomaly''). 
More recently, intense investigation of the phase behavior of potentials
with a softened core has shown that these models can yield, even for
isotropic one-component systems, unusual properties such as a melting
line with a maximum followed by a region of re-entrant melting,
polymorphism both in the fluid and solid phases, and a rich hierarchy of
water-like anomalies~\cite{Stillinger,Sadr,Jagla,Kumar,Xu,Deoliveira1,
Gibson,Lomba,Fomin,Deoliveira2,Pizio,Pamies,Pauschenwein}.

The common feature of all the softened-core potentials investigated so far
is that, in a range of interparticle distances $r$, the strength of the
two-body force $f(r)\equiv-u'(r)$ reduces or at most remains constant as
two particles approach each other, $u(r)$ being the repulsive part of the
pair potential.
In this interval of distances, $u(r)$ shows a downward or zero concavity,
i.e., $u''(r)\le 0$. Assuming that the repulsion is hard-core-like
at small distances and goes to zero sufficiently fast at large distances,
the above behavior makes it possible to identify two distinct regions where
the repulsive force increases as $r$ gets smaller: a inner region, associated
with the impenetrable particle core, and an outer region at large distances.
Such potentials are thus endowed with two repulsive length scales:
a smaller one (``hard'' radius), which is dominant at the higher pressures,
and a larger one (``soft'' radius), being effective at low pressure.
In the range of
pressures where the two length scales compete, a system governed by soft-core
interactions behaves as a ``two-state'' system, a simplified viewpoint that
nonetheless provides an explanation for re-entrant melting and, in presence
of an additional attractive long-range force, for the existence of a
liquid-liquid transition~\cite{Mcmillan}.

Though core-softening has been usually associated to a repulsive force 
with non-monotonic behavior, the latter condition might be an unnecessary
requirement. This observation stems from an analysis of the mathematical
expression of core-softening, which was put forward by Debenedetti et
al.~\cite{Debenedetti} under the form: $\Delta(rf(r))<0$ for $\Delta r<0$,
in some interval $r_1<r<r_2$, together with $u''(r)>0$ for $r<r_1$ and $r>r_2$.
The above conditions are satisfied if, in the interval $(r_1,r_2)$,
the product $rf(r)$ (rather than just $f$) reduces with decreasing
interparticle separation.
Hence, the core-softening property may also be satisfied with a strictly
convex potential, yielding a repulsive force which everywhere increases for
decreasing $r$, provided that, in a range of interparticle distances, the
increasing rate of $f(r)$ be sufficiently small. On this basis, we expect
that the class of core-softened potentials is in fact much wider
than thought before.

We hereafter study the phase diagram of a model potential which is soft,
according to Debenedetti's formulation, though being characterized by a
{\em strictly monotonic} force for all distances (i.e., $u''(r)>0$ everywhere).
We find that this potential exhibits the whole spectrum of anomalies
that are usually associated with soft-core potentials,
including re-entrant melting regions, solid polymorphism, and water-like
anomalies. This paper is organized as follows: in Sec.\,2, we describe the
model system which is the subject of our investigation. In Sec.\,3,
we outline the simulation method that is used to map out the phase
diagram. The simulation results are presented and discussed in Sec.\,4
while further remarks and conclusions are postponed to Sec.\,5.

\section{Model}

We consider a purely repulsive pair potential modelled through an
exponential form, first introduced, about four decades ago, by Yoshida
and Kamakura (YK)~\cite{Yoshida}:
\be
u(r)=\epsilon\exp\left\{a\left(1-\frac{r}{\sigma}\right)-6
\left(1-\frac{r}{\sigma}\right)^2\ln\left(\frac{r}{\sigma}\right)\right\}\,,
\label{eq1}
\ee
where $\epsilon$ and $\sigma$ are the energy and length units respectively
and $a>0$. This potential behaves as $r^{-6}$ for small $r$, and falls
off very rapidly for large $r$. The softness of the repulsion is controlled 
by the $a$ parameter. When $a<1.9$, the YK potential has a region with
downward curvature where the force decreases as two particles get closer.

Approximate theoretical calculations~\cite{Yoshida} suggest that the melting
line of the YK potential might display a re-entrant melting region for values
of $a$ that are larger than 1.9, i.e., even when no downward concavity is
present. In order to explore this possibility and discuss it in relation with 
other anomalous behaviors, we have investigated the phase behavior of the
YK potential for $a=2.1$ through numerical simulation. 
For this $a$ value (as well as for all $a$'s falling in the range $1.9-2.3$) $u(r)$ is strictly convex (see Fig.\,1), i.e., its second derivative is positive everywhere, hence the repulsive force is strictly increasing for decreasing $r$, at variance with the core-softened potentials considered so far. However, the rate at which the force increases is not monotonous. By the way, in a range of $r$ that roughly corresponds to the local minimum of $u''(r)$, the repulsive force increases with
decreasing $r$ much more slowly than in the adjacent regions, in such a
way that the Debenedetti core-softening property is still satisfied.

\section{Method}

A major problem, when drawing the phase diagram of a model system
of particles interacting through an assigned potential, is to identify the
solid phases. This is a critical issue for soft-core repulsions since
experience has shown that many different crystal structures are
stabilized for such systems upon varying the
pressure~\cite{Likos3,Fomin,Pamies,Prestipino}. In the present study,
we consider as candidates for the solid phase precisely those crystal
structures that a previous investigation of the same
potential~\cite{Prestipino} showed to be stable at zero temperature.
These include the face-centered cubic (fcc), body-centered cubic (bcc), and
simple hexagonal (sh) lattices, as well as a number of non-Bravais lattices,
i.e., the A7, diamond, A20, and hexagonal close-packed (hcp) lattices.
On increasing pressure, each of these structures gives, in turn, the most
stable structure at $T=0$. The sequence of stable crystals for increasing
pressures was found to be~\cite{Prestipino}:
\be
{\rm fcc}\stackrel{0.63}{\longrightarrow}{\rm bcc}
\stackrel{1.26}{\longrightarrow}{\rm sh}
\stackrel{2.29}{\longrightarrow}{\rm A7}
\stackrel{2.55}{\longrightarrow}{\rm diamond}
\stackrel{4.91}{\longrightarrow}{\rm sh}
\stackrel{5.46}{\longrightarrow}{\rm A20}
\stackrel{12.07}{\longrightarrow}{\rm hcp}
\stackrel{15.68}{\longrightarrow}{\rm bcc}
\stackrel{52.75}{\longrightarrow}{\rm fcc}
\stackrel{138.28}{\longrightarrow}{\rm hcp}
\stackrel{365.65}{\longrightarrow}{\rm fcc}\,,
\label{eq2}
\ee
where the numbers above the arrows indicate the transition pressures
expressed, to within a precision of 0.01, in units of $\epsilon/\sigma^3$.
Where pertinent, the values of the internal parameters of the phases
listed in Eq.\,\ref{eq2} are reported in Ref.\,\cite{Prestipino}.
Although the above sequence of phases resulted from a careful scrutiny of
about thirty crystal structures, we cannot exclude that some relevant
structure might have been overlooked.
In general, we may expect that the same crystals that are stable at $T=0$
also give the underlying lattice structure for the solid phases at $T>0$.
Anyway, if more crystals exist at $T=0$ with nearly the same chemical
potential $\mu$ in a pressure range, each of them represents a potentially
relevant solid phase. In our case, this occurs for
$P\approx 2.4$, where the ground state is of A7 type but some oC8 and cI16
crystals have only slightly larger chemical potentials. Hence, we include
also these lattices in our list of solid candidates.

We calculate the phase diagram of the YK model for $a=2.1$ by performing
Monte Carlo (MC) simulations in the isothermal-isobaric, $NPT$ ensemble
(i.e., at constant temperature $T$, pressure $P$ and number $N$ of particles), 
as well as in the canonical, $NVT$ ensemble, using the standard Metropolis
algorithm with periodic boundary conditions and the nearest-image convention.
In the solid phases, the number of particles is chosen so as to guarantee a
negligible contribution to the interaction energy from pairs of particles
separated by half the minimum simulation-box length. With this choice, we
checked in a number of cases that the exact location of phase boundaries
is only negligibly affected by the finite size of the sample. More precisely,
our samples consisted of $500$ particles in the fcc phase, of $432$ particles
in the bcc phase, of $800$ particles in the sh phase, of $512$ particles
in the diamond phase, of $1024$ particles in the cI16 phase, of $1152$
particles in the A7 phase, and of $768$ particles in the oC8 phase.
Depending on the solid phase with which the fluid is compared to in order
to assess its relative stability, we consider fluid samples of $500$ to
$800$ particles. At given $T$ and $P$, sample equilibration
typically took from 20000 to 50000 MC sweeps, a sweep consisting of one
average attempt per particle to change its position plus (for the $NPT$ case
only) one attempt to change the volume through a rescaling of particle
coordinates.
The maximum random displacement of a particle and the maximum volume change
in a trial MC move are adjusted every sweep during the equilibration run so
as to obtain an acceptance ratio of moves close to 50\% (afterwards, during
the production session, they are maintained fixed). Thermodynamic averages
are computed over trajectories from $3\times 10^4$ to $10^5$ sweeps long.

In a typical $NPT$ simulation, we generate a sequence of simulation runs
along an isobar, starting from the cold solid on one side and from the hot
fluid on the other side (a chain of $NVT$ runs along an isotherm at a
sufficiently high temperature provides the link to a dilute-fluid state).
The last configuration produced in a given run is taken to be the first
of the next run at a slightly different temperature. The starting
configuration of a ``solid'' chain of runs is a low-temperature perfect
crystal whose density is set equal to its $T=0$ value at that pressure.
In case of a structure with internal parameters, the same $T=0$ optimal
parameters are chosen in the preparation of the crystal sample.
Usually, a series of runs is continued until a sudden change is
found in the difference of energy/volume between the solid and the
fluid, so as to avoid averaging over heterogeneous thermodynamic states.
The density of a solid phase ordinarily varies very little with increasing
temperature along an isobar. A sudden density change thus indicates
a mechanic instability of the solid in favour of the fluid, hence it
signals the approximate location of melting. In fact, this so called
``heat-until-it-melts'' (HUIM) procedure allows to locate the maximum
overheating temperature $T'_m$, which may however be substantially
higher than the fluid-solid coexistence temperature $T_m$.

In order to compute the difference in chemical potential between any two
equilibrium states of the system (say, 1 and 2) belonging to the same phase,
we use the standard thermodynamic-integration method. This allows to obtain
a thermodynamic potential as an integral over the simulation path of a
calculated statistical average (energy, density, or pressure, depending on
the path followed). To be specific, we use
\be
f_{\rm ex}(T,\rho_2)-f_{\rm ex}(T,\rho_1)=
k_BT\int_{\rho_1}^{\rho_2}\frac{{\rm d}\rho}{\rho}
\left(\frac{P(T,\rho)}{\rho k_BT}-1\right)
\label{eq3}
\ee
along an isothermal $NVT$ simulation path, $f_{\rm ex}$ being the excess
Helmholtz free energy, while we use the formulae
\be
\mu(T,P_2)-\mu(T,P_1)=\int_{P_1}^{P_2}{\rm d}P\,v(T,P)
\label{eq4}
\ee
and
\be
\frac{\mu(T_2,P)}{T_2}-\frac{\mu(T_1,P)}{T_1}=
-\int_{T_1}^{T_2}{\rm d}T\,\frac{u(T,P)+Pv(T,P)}{T^2}
\label{eq5}
\ee
along an isothermal and isobaric $NPT$ path, respectively, $u$ and $v$
being the specific values of energy and volume.
To prove useful, the above equations require an independent estimate
of $f_{\rm ex}$ or $\mu$ in one reference state for each phase. For the fluid,
a reference state can be any state being characterized by a very small system
density, since then the excess chemical potential can be accurately estimated
through Widom's particle-insertion method~\cite{Widom}. In order to calculate
the excess Helmholtz free energy of a solid, we resort to the Frenkel-Ladd
method~\cite{Frenkel1} (see Ref.\,\cite{Prestipino2} for a full description
of the method and of its implementation on a computer). The solid excess
Helmholtz free energy is calculated through a series of $NVT$ simulation
runs at a fixed state point, {\it i.e.}, for fixed density and temperature.

\section{Results and discussion}

We arbitrarily restrict our analysis of the phase diagram to pressures
$P$ smaller than $5$ (in reduced units), which corresponds approximately to the upper limit of stability of the diamond crystal at $T=0$~\cite{Prestipino}.
For a number of pressures, we first calculate the fluid-solid coexistence
temperature $T_m$ by employing the ``exact'' free-energy method described
above. By estimating the maximum overheating temperature $T'_m$ through the
HUIM method, we find that the difference between $T'_m(P)$ and $T_m(P)$
is always small in relative terms (less than 15\%) and is significant
only at the highest pressures $P$. Hence, we use the more rapid HUIM
approach to derive the overall trend of the melting line. The investigated
system shows a rich solid polymorphism at $T>0$ (see Fig.\,2), which is
closely related to the alternation of crystal phases at zero temperature.
In the low-pressure region ($P<1$), upon increasing
pressure at low temperatures ($T<0.04$, in $\epsilon/k_B$ units) the
fluid freezes first into a fcc solid, which then undergoes a transition
into a bcc solid.
At higher temperatures, in a narrow $T$ interval, the sequence of phase 
transitions undergone by our system with increasing pressures is
fluid-bcc-fcc-bcc. This behavior is similar to that observed for the
Gaussian core model~\cite{Prestipino3} and occurs from $T_{\rm tr}$ to
$T\simeq 0.05$, $T_{\rm tr}\lesssim 0.045$ being the fluid-bcc-fcc
triple-point temperature.
For $0.05<T<0.06$, the fcc phase ceases to be stable and the fluid freezes
directly into a bcc solid. The bcc-fluid coexistence line shows, at
$P\simeq 0.6$, a maximum melting temperature; above this pressure, the
${\rm d}T/{\rm d}P$ slope of the bcc melting line is negative.
Thus, the bcc solid undergoes, for not too low temperatures, re-entrant
melting into a denser fluid.

At low temperatures, the bcc solid transforms, for pressures around
$1.3$, into a solid with a sh structure. Hence, there should be a
fluid-bcc-sh triple point which terminates the bcc reentrant melting
line. This is illustrated in detail in Fig.\,3, where we show the
calculated chemical potentials of the three phases along the $T=0.03$
isotherm. Upon further increasing pressure, a $cI16$ solid with internal
parameter $x=0.125$ becomes stable (see Ref.\,\cite{Prestipino} for
a definition of $x$). This can occur only for high enough $T$ since,
in the pressure interval $2.30<P<2.55$, a certain A7 crystal has a
lower chemical potential than cI16 at $T=0$ (we have checked that no
oC8 solid is stable in the same temperature range). In other words, a
small A7 basin exists below $T\approx 0.01$. Following the cI16 phase,
another solid phase arranged according to the diamond structure becomes
stable in a broad pressure range, starting from $P\simeq 2.6$.

Within the precision of our calculation, the overall melting line exhibits,
in addition to the first (bcc-fluid) re-entrant region, other portions with
negative ${\rm d}T/{\rm d}P$ slope, though less pronounced than the first.
The alternation of solid phases at low temperatures goes on beyond $P=5$
until, eventually, the system sets in a fcc solid that coexists with the
fluid phase at arbitrary high temperatures, the coexistence line becoming
asymptotically the same as for the $r^{-6}$ potential. The complex phase
behavior shown in Fig.\,2 is absent in the theoretical calculation of
Yoshida {\it et al.}~\cite{Yoshida}, where only the possibility of a fcc
solid was taken into account. With this limitation, only one region of
reentrant melting is predicted and the high-pressure fluid can be stable
even at zero temperature.
Moreover, the height and width of the low-density solid region are largely
overestimated by the theory as compared to the simulation results here
presented.

Our results show that the model system here considered, though described
by a spherically-symmetric potential, can exist under the form of stable
non-Bravais crystals. The rich polymorphism observed follows from the
peculiar dependence of the interatomic force with distance, which leads
to the existence of two incommensurate length scales. In the pressure
and temperature regimes where these lengths compete with each other,
compact arrangements such as the fcc and bcc lattices are frustrated and
low-coordinated arrangements are preferred. Our results are relevant for
those physical systems that are characterized by a certain softness of
the repulsive interaction. These include intrinsically soft materials
such as colloids, polymers, etc. but also atomic systems at extremely
high pressures. Phase diagrams with solid
polymorphism and multiple re-entrant melting have been observed at high
pressures in some elements, such as Sr~\cite{Errandonea} and predicted
by ab-initio simulation for others~\cite{Correa}.
Concerning model systems, a phase behavior with features similar to those
noted above was found for the hard-core plus repulsive step
potential~\cite{Fomin} and for Hertzian spheres~\cite{Pamies}.

Some insight into the mechanisms that give origin to the complex phase
behavior observed can be got from analysing the radial distribution function
$g(r)$.
We computed $g(r)$ for various pressure values, at a temperature slightly
larger than the maximum melting temperature $T_M\simeq 0.06$ of the bcc solid
(Fig.\,4).
At very low pressure, the first peak of $g(r)$ is centred around
$r\lesssim 2$ (in units of $\sigma$). As $P$ increases up to $P\approx 0.50$,
this peak moves upward
while its position shifts towards $r\simeq 1.5$. For further pressure
increase, its height decreases while its position remains almost unaltered.
At the same time, a new peak develops around $r=1$, whose height grows with
$P$. This behavior is significantly different from that of simple fluids,
where all peaks of $g(r)$ get higher with pressure when keeping $T$ constant,
and is consistent with the gradual penetration
of particles inside the inner core. Thus, the analysis of $g(r)$ points to
the existence of two competitive scales of nearest-neighbour distance,
i.e., a larger soft length fading out with increasing pressure in favour of the smaller hard length.
The soft-repulsive distance falls approximately at $r=1.5$, which is
where the second derivative of the potential has a local maximum,
while the hard length scale remains defined by the extremely rapid
increase of the short-range repulsive force around $r=1$.

Thermodynamic, dynamic and structural anomalies have been observed in a
number of substances (such as water, silica, silicon, carbon, and
phosphorous)~\cite{Mcmillan}. These unconventional features are usually
referred to as water-like anomalies. Although these substances are
all characterized by local tetrahedral order, in the last years water-like
anomalies have been found also in systems with spherically symmetric
potentials, provided that the repulsion is of the softened-core
type~\cite{Stillinger,Sadr,Jagla,Kumar,Xu,Deoliveira1,Gibson,Lomba,Fomin,
Deoliveira2,Pizio,Pamies,Pauschenwein}
In order to investigate the existence of
anomalies in our system, we first analysed the behavior of the number
density in the fluid phase above the melting line. We find that in the
region lying approximately above the re-entrant portion of the bcc-fluid
coexistence line, decreasing temperature at constant pressure leads first
to a density increase and then, contrary to standard behavior, to a density
decrease for further cooling (Fig.\,5). The anomalous behavior of the density
can be interpreted in terms of the existence of two repulsive
length scales: as $T$ reduces at constant pressure, the larger soft scale
becomes favoured with respect to the smaller hard one.
Thus, compact local structures are less favoured than open ones, which
causes a decrease in the number of particles within a given volume. 
The $P$-$T$ region where the density behaves anomalously is bounded from
the above by the locus of points where the density attains its maximum
value, i.e., by the temperature of density maximum (TMD) line, while its
lower bound is the limit of stability of the fluid phase, namely the bcc-fluid
coexistence line (see Fig.\,2).
Within the region of density anomaly, the system expands upon cooling under
fixed pressure. Consistently, the thermal expansion coefficient
$\alpha_P=(1/V)(\partial V/\partial T)_P$ is negative,
vanishing along the TMD line (Fig.\,5).

A series of studies have shown that thermodynamic (and dynamic) anomalies
are strongly correlated with anomalous trends in the structural order of
the system~\cite{Errington}. A measure of this quantity is provided by the
pair entropy per particle,
\be
s_2=-\frac{k_B}{2}\rho\int{\rm d}^3r\,\left[g(r)\ln g(r)-g(r)+1\right]\,,
\label{s2}
\ee
which describes the contribution of two-body spatial correlations to
the excess entropy of the fluid~\cite{Nettleton}.
For a completely uncorrelated system, $g(r)=1$ and then $s_2=0$.
For systems with long-range order, spatial correlations persist
over large distances and $s_2$ becomes large and negative.
Thus $-s_2$ can be taken as an indicator of structural order.
This parameter yields information about the average relative spacing of
the particles, i.e., it describes the tendency of particle pairs to adopt
preferential separations.
In the investigated system, the structural order initially increases upon
compression (Fig.\,6), similarly to simple fluids. However, as pressure is
further increased, $-s_2$ attains a maximum value and then decreases,
i.e., the fluid loses structural order upon compression (structural anomaly). 
At sufficiently high density, after attaining a local minimum,
$-s_2$ recovers a normal trend, increasing monotonously with pressure.
Upon increasing the temperature, the structural anomaly becomes less and
less marked (i.e., the difference between the local extrema of $-s_2$ gets
smaller).
As seen in Fig.\,2, the region where the fluid has a structurally anomalous
behavior embraces the density anomaly region; a similar relationship between
the structural and density anomalies holds for water and for a number of model
systems~\cite{Errington,Esposito,Yan,Yan2}.

Though, from a thermodynamic point of view, phase coexistence is determined
solely by the equality of the Gibbs free energy in the two phases, an
incoming phase transition may induce a number of modifications in the
involved phases (the analysis of such changes provides the basis for
the so-called ``one-phase'' criteria for phase transitions). Here, we
investigate to what extent the rich solid polymorphism shown by our system
at low temperature is reflected in the fluid lying at higher temperatures.
The structural order, as measured by $-s_2$, offers interesting insight about
this point. The behavior of $-s_2$ shows that the low-density bcc
region casts an imprint on the surrounding fluid, under the form of
a marked increase of the structural order near the pressure of maximum
melting temperature of the solid. On the contrary, the detailed shape
of the melting curve at intermediate densities does not significantly affect
the structural order of the fluid, which shows a steady enhancement
with pressure. In order to observe an appreciable modification of this
behavior it is necessary to examine the fluid very close to the solid.
Only the lowest $-s_2$ isotherm at $T=0.05$, i.e., just at the lower edge
of the fluid phase, shows a modest bump reflecting the fine details of
the melting line (see Fig.\,6).
These results suggest that the influence of low-coordinated solid structures
on the structural order of the neighbouring fluid is much weaker than for
compact lattices. From the analysis of the $g(r)$, it clearly appears that
the soft-repulsive scale looses efficacy
in the pressure range corresponding approximately to the re-entrant region of
the bcc solid, where both density and structural anomalies occur.
Beyond $P=1.5$, the second peak of $g(r)$ changes only slightly (decreasing
with increasing density), while the first peak builds up significantly with
pressure, which is the main reason for the steady increase of
$-s_2$ at intermediate pressures (this is related to the gradual increase
of local order with compression, until the fluid crystallizes into a
fcc structure at very high pressures, out of the range shown in Fig.\,2).
The third peak undergoes, with increasing pressure, subtle but
significant changes, splitting in two minor peaks (Fig.\,7). Such changes
mirror the alternation of the solid structures at low $T$ but, being small
modifications of the $g(r)$, they scarcely affect the structural order of
the fluid, at least that quantified by $-s_2$.

\section{Concluding remarks}

The occurrence of anomalous phase behavior within the class of isotropic
one-component classical potentials was, up to now, thought to be possible
only for potentials with a region of downward or zero curvature, where the
repulsive force decreases or is at most constant as two particles get closer.
Here, we have shown that anomalous phase behaviors can actually occur also
for a system of particles interacting through a strictly monotonic repulsive
force, provided that in a range of interparticle distances the force increases
more slowly, with decreasing $r$, than in the adjacent regions.
This condition gives origin, in spite of the convexity of the potential,
to two distinct repulsive length scales, a feature that seems instrumental
for the occurrence of re-entrant melting and the related water-like anomalies.
From the present results, we may expect that the real systems effectively
characterized by isotropic interactions and able to show unusual phase
behaviors are more numerous than previously believed.

\clearpage


\begin{figure}
\includegraphics[width=12cm,angle=0]{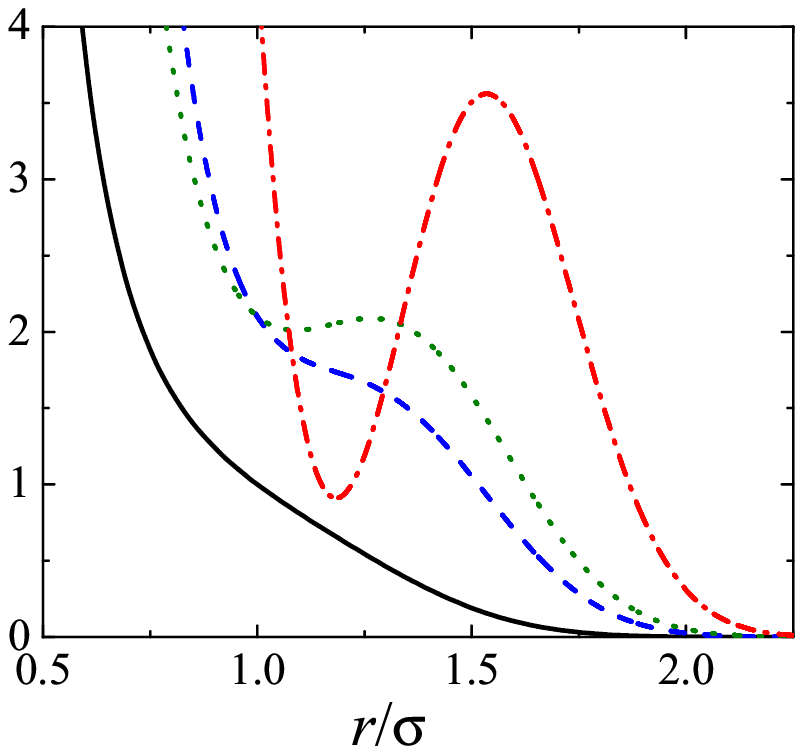}
\caption{(Color online). Yoshida-Kamakura potential $u(r)$ for $a=2.1$
(black solid line, expressed in $\epsilon$ units), two-body force
$f(r)=-u'(r)$ (blue dashed line, $\epsilon/\sigma$ units), product
$rf(r)$ (green dotted line, $\epsilon$ units), second derivative of
the potential $u''(r)$ (red dash-dotted line, $\epsilon/\sigma^{2}$ units)}
\label{fig1}
\end{figure}

\begin{figure}
\includegraphics[width=12cm,angle=0]{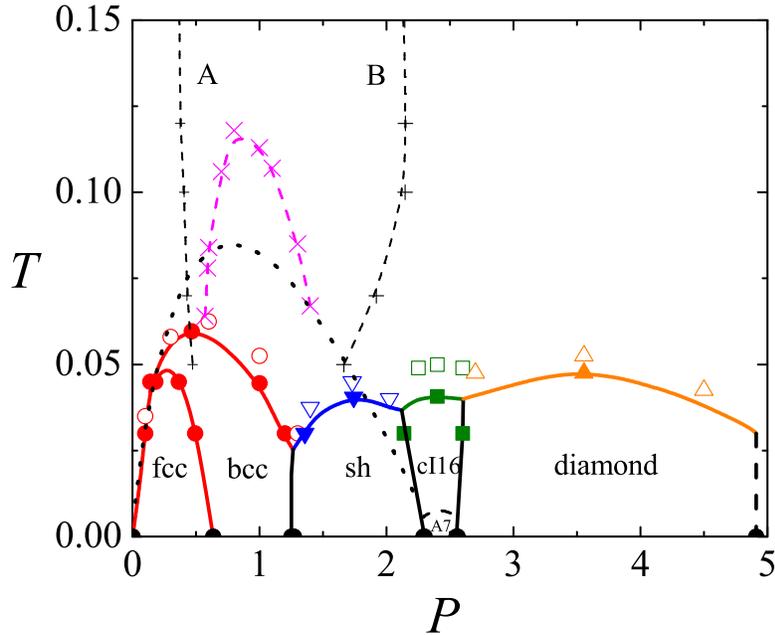}
\caption{(Color online). Phase diagram of the Yoshida-Kamakura interaction
model ($a=2.1$). Pressure $P$ and temperature $T$ are expressed in units of
$\epsilon/\sigma^3$ and $\epsilon/k_B$, respectively ($k_B$ being Boltzmann's
constant). The phase-transition points, obtained by exact free-energy
calculations, are represented as full symbols (different colours refer to
different solid structures; errors are smaller than the size of the symbols).
The data poins lying on the $T=0$ axis are exact solid-solid
boundaries~\cite{Prestipino}.
The solid lines through the transition points are tentative phase boundaries,
drawn following the trend of thresholds of maximum solid overheating,
hereby reported as open dots. The location of the dashed line delimiting
the A7 region from the above is uncertain, since we did not carry out any
exact free-energy calculation for the A7 solid.
The dashed line connecting crosses is the locus of density maxima in the
fluid phase. Curves A and B connect points of maximum and minimum values of
$-s_{2}$,
respectively (see Fig.\,6). The open region between A and B is the structurally
anomalous region. The black dotted line is the melting line as estimated in
\cite{Yoshida}. All lines in the figure are guides for the eye.}
\label{fig2}
\end{figure}

\begin{figure}
\includegraphics[width=12cm,angle=0]{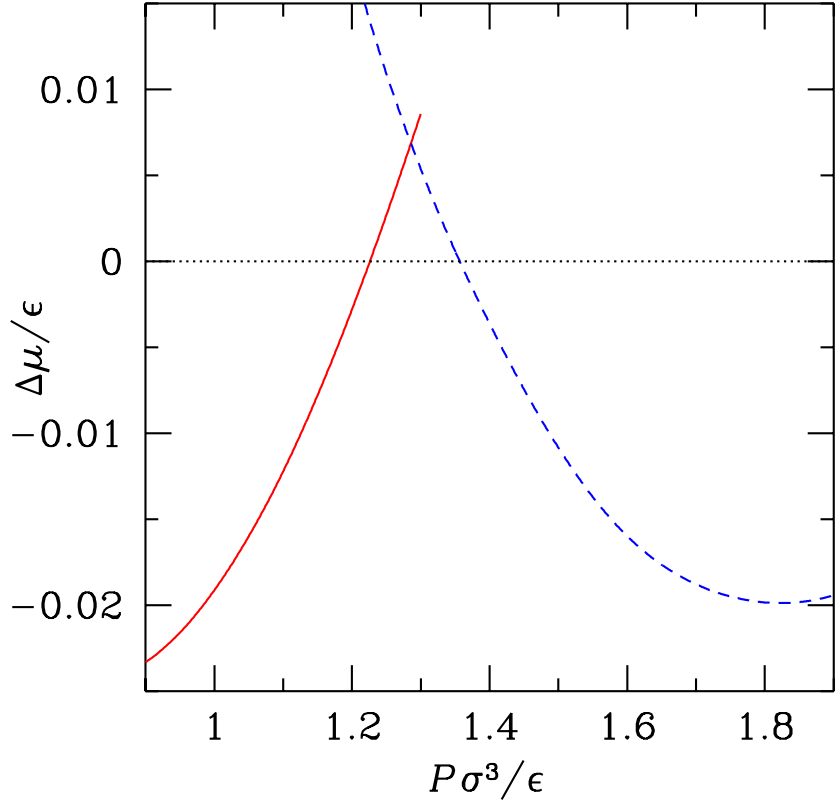}
\caption{(Color online). Chemical potential $\mu$ of the bcc and sh solids at
$T=0.03$,
taking the fluid phase as reference: bcc (solid red line), sh (dashed blue
line), and fluid (dotted line). The stable phase is the one with lower
$\mu$, hence the equilibrium system is fluid between 1.226 (bcc-fluid
transition pressure) and 1.357 (fluid-sh transition pressure). Note
the existence of a minimum in the $\mu$ difference between sh and fluid
at $P\simeq 1.8$, which roughly corresponds to the pressure
at which the sh melting temperature is maximum.}
\label{fig3}
\end{figure}

\begin{figure}
\includegraphics[width=12cm,angle=0]{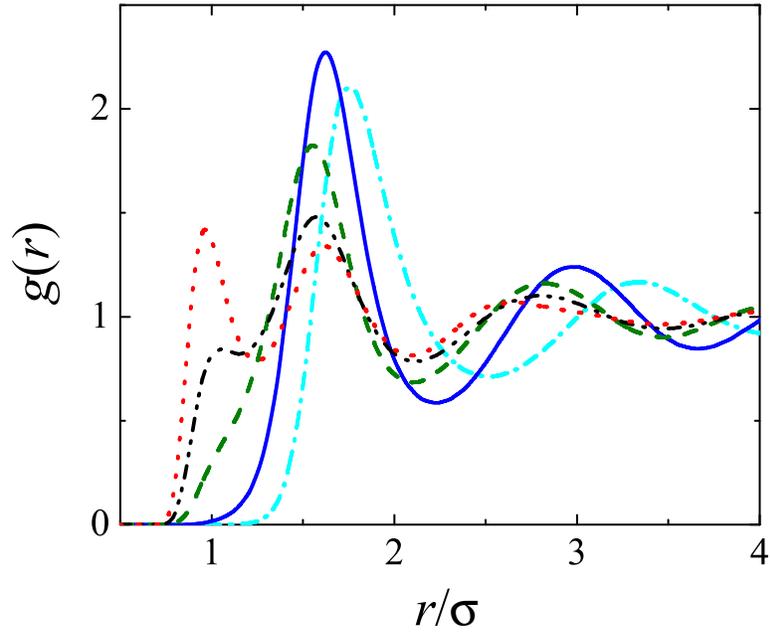}
\caption{(Color online). Pair distribution function $g(r)$ for $T = 0.07$:
$P=0.15$ (cyan dash-dotted line); $P=0.49$ (blue solid line);
$P=0.99$ (green dashed line); $P=1.56$
(black dash-dot-dotted line); $P=3.02$ (red dotted line).}
\label{fig4}
\end{figure}

\begin{figure}
\includegraphics[width=12cm,angle=0]{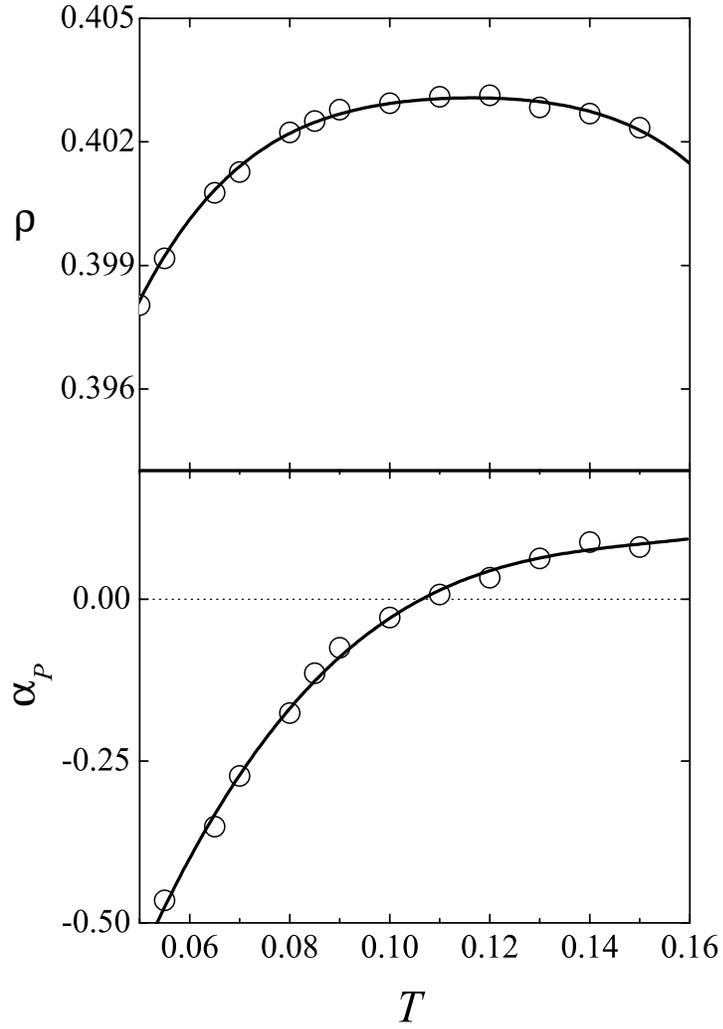}
\caption{Reduced number density $\rho$ (upper panel) and thermal-expansion
coefficient $\alpha_P$ (lower panel) plotted as a function of temperature
for $P=1$. The solid lines are polynomial fits of the data. Note that
$\alpha_P$ vanishes approximately where $\rho$ is maximum.}
\label{fig5}
\end{figure}

\begin{figure}
\includegraphics[width=12cm,angle=0]{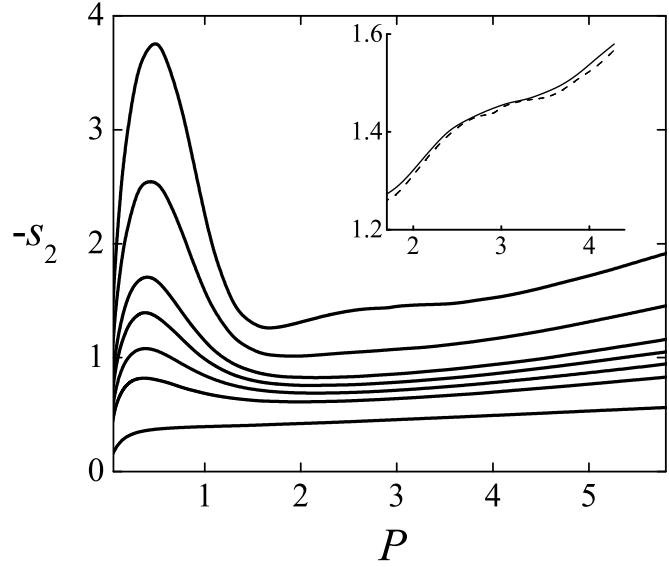}
\caption{Structural order parameter $-s_2$, plotted for
$N=500$ as a function of the reduced pressure at different temperatures:
from top to bottom, $T=0.05,0.07,0.1,0.12,0.15,0.2$ and 0.50.
The inset shows a magnification of $-s_2$ for $T=0.05$
(dashed line: $N=500$; solid line: $N=800$)}
\label{fig6}
\end{figure}

\begin{figure}
\includegraphics[width=12cm,angle=0]{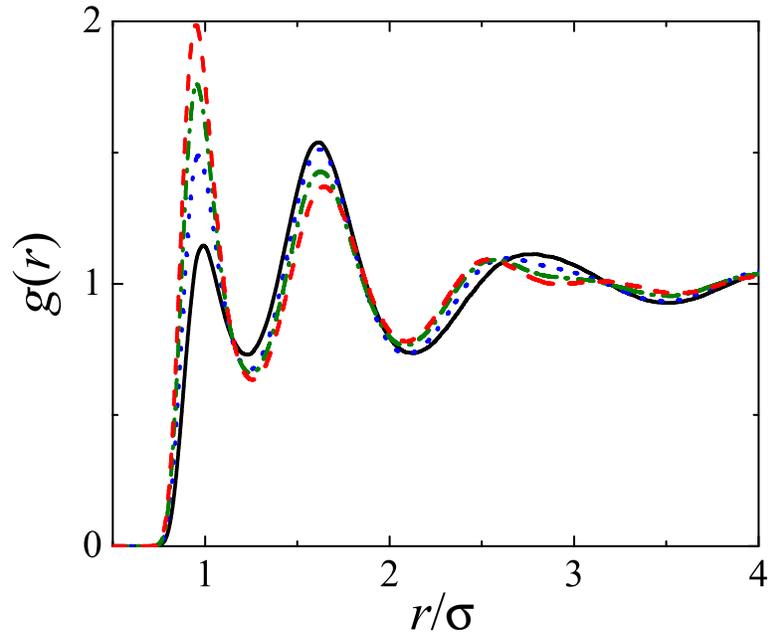}
\caption{(Color online). Pair distribution function $g(r)$ for $T=0.05$:
$P=1.95$ (black solid line); $P=2.75$ (blue dotted line); $P=3.67$
(green dash-dotted line); $P=4.71$ (red dashed line).}
\label{fig7}
\end{figure}
\end{document}